# A symmetric formula of transformed elasticity tensor in PML domain for elastic wave problem

Yingshi Chen[1]

[1] Institute of Electromagnetics and Acoustics, and Department of Electronic Science, Xiamen University, Xiamen 361005, China

E-mail: gsp@grusoft.com

**Abstract**

PML(Perfectly matched layer) is very important for the elastic wave problem in the frequency domain. Generally, the formulas of elasticity tensor in PML region are derived from the transformed momentum equation. In this note, we proved that the transformed elasticity tensor derived in this way lost its symmetry. Therefore, these formulas are inconsistency in theory and it's hard to explain its numerical performance. We present a new symmetrical formula of elasticity tensor from the weak form. So the theory of elasticity is still applicable in PML domain.

PML(Perfectly matched layer) is very important for the elastic wave problem in the frequency domain. The formulas of many papers are derived from the transformed momentum equation. However, in this way, the transformed elasticity tensor lost its symmetry. Therefore, in theory, these formulas are inconsistency and it's hard to explain its numerical performance [5]. In this note, we get a new symmetrical formula of elasticity tensor from the weak form.

We use Cartesian tensors such as $\tau_{ji}$, where the indices $i,j$=1,2,3. We also use Einstein Summation Convention: if subscripted variables appearing twice in any term, the subscripted variables are assumed to be summed over.

Along each coordinate axis, we define the unit orthonormal base vector $\mathbf{e}_i$.

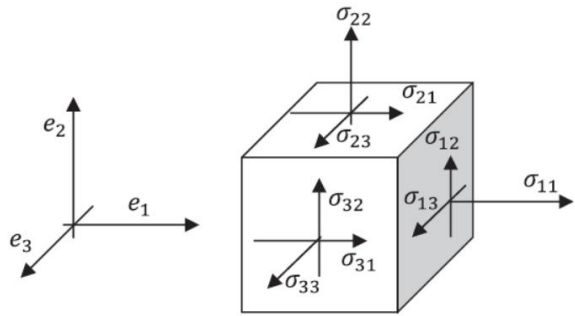

From Euler's momentum equation [1, 2], we get elastic wave equation in the frequency domain.

$$-\omega^2 \rho u_i - \frac{\partial \tau_{ji}}{\partial x_j} = f_i \qquad (1)$$

$$\text{or} \quad -\omega^2 \rho u_i - \frac{\partial\left(C_{jikl} \cdot \frac{\partial u_k}{\partial x_l}\right)}{\partial x_j} = f_i \quad (1.1)$$

(Note: *the expand of $\tau_{ji}$ in 1.1 need the symmetry of $C_{jikl}$*)

To absorb wave in the PML region, we use the famous complex coordinate stretching technique [6,7,8].

## Unsymmetrical formula from transformed momentum equation

Many papers [9,3] try to get transformed momentum equation, which includes some new mathematic objects $\tilde{\rho}, \tilde{\tau}_{ij}, \tilde{C}_{ijkl}$.

$$-\omega^2 \tilde{\rho} u_i - \frac{\partial \tilde{\tau}_{ji}}{\partial x_j} = \tilde{f}_i \quad (2)$$

where

$$\tilde{\rho} = \rho s_1 s_2 s_3$$

$$\tilde{\tau}_{ij} = \tilde{C}_{ijkl} \frac{\partial u_l}{\partial x_k} \quad (3)$$

$$\tilde{C}_{ijkl} = C_{ijkl} \frac{s_1 s_2 s_3}{s_i s_k} \quad (4)$$

It's easy to verify $\tilde{C}_{ijkl} \neq \tilde{C}_{jikl}$ (e.g. $\tilde{C}_{12kl} = C_{12kl} \frac{s_2 s_3}{s_k}, \tilde{C}_{21kl} = C_{21kl} \frac{s_1 s_3}{s_k}$), then

$$\tilde{\tau}_{ij} \neq \tilde{\tau}_{ji} \quad (5)$$

The asymmetry of $\tilde{\tau}_{ij}$ means that theory of elasticity breaks down in the PML domain! Then any formula based on (2) are suspicious. As [5] showed: "if the stretched stresses fail equilibrium, then none of the classical theorems of elasticity which underpin the finite element method —for example, the principle of virtual work— can be assumed without much ado to remain strictly valid for PMLs."

So this method is unconvincing and hard to explain its numerical performance

## Symmetrical formula from weak form

Apply the standard Galerkin procedure, we get weak form

$$-\omega^2 \int_V \emptyset \rho u_i \, dv - \int_V \emptyset \frac{\partial \tau_{ji}}{\partial x_j} dv = \int_V \emptyset f_i \, dv \quad (6)$$

where $\emptyset$ is any scalar testing function。

After integration by parts

$$-\omega^2 \int_V \emptyset \rho u_i \, dv + \int_V \frac{\partial \emptyset}{\partial x_j} \tau_{ji} \, dv - \int_V \frac{\partial(\emptyset \tau_{ji})}{\partial x_j} dv = \int_V \emptyset f_i \, dv \quad (7)$$

Let's check the third term, which includes second order of the displacement vector **u**. By gauss (divergence) theorem:

$$\int_V \frac{\partial(\emptyset \tau_{ji})}{\partial x_j} dv = \int_V \left(\frac{\partial(\emptyset \tau_{1i})}{\partial x_1} + \frac{\partial(\emptyset \tau_{2i})}{\partial x_2} + \frac{\partial(\emptyset \tau_{3i})}{\partial x_3}\right) dv = \int_S \emptyset (\tau_{1i} n_1 + \tau_{2i} n_2 + \tau_{3i} n_3) \, ds \quad (8)$$

In the linear elastic theory, at **free surface**, $\tau_{ji} = \tau_{ij}=0$, So this term is zero. Or use zero dirichlet boundary condition, $\emptyset = 0$ in the surface, this term is also 0.

Let' check the second term $\int_V \frac{\partial \emptyset}{\partial x_j} \tau_{ji}\, dv$. For example, let i=1

$$\int_V \frac{\partial \emptyset}{\partial x_j} \tau_{j1} dv = \int_E \frac{\partial \emptyset}{\partial x_j}\left(C_{j1kl} : \frac{\partial u_k}{\partial x_l}\right) dv =$$

$$\sum_{j=1}^{3} \frac{\partial \emptyset}{\partial x_j}\left(C_{j1kl} : \frac{\partial u_k}{\partial x_l}\right) = \sum_{j=1}^{3} \frac{\partial \emptyset}{\partial x_j}(C_{1j1}\nabla u_1 + C_{1j2}\nabla u_2 + C_{1j3}\nabla u_3)$$

$$= \nabla\emptyset \cdot (C_{11}\nabla u_1) + \nabla\emptyset \cdot (C_{12}\nabla u_2) + \nabla\emptyset \cdot (C_{13}\nabla u_3) \quad (9)$$

or $\sum_{j=1}^{3} \frac{\partial \emptyset}{\partial x_j}\left(C_{j1kl} : \frac{\partial u_k}{\partial x_l}\right) = \nabla\emptyset \cdot [C_{11}\nabla u_1 + C_{12}\nabla u_2 + C_{13}\nabla u_3]$ (10)

where

$$C_{11} = \begin{pmatrix} c_{1111} & c_{1112} & c_{1113} \\ c_{2111} & c_{2112} & c_{2113} \\ c_{3111} & c_{3112} & c_{3113} \end{pmatrix} \quad C_{12} = \begin{pmatrix} c_{1121} & c_{1122} & c_{1123} \\ c_{2121} & c_{2122} & c_{2123} \\ c_{3121} & c_{3122} & c_{3123} \end{pmatrix} \quad C_{13} = \begin{pmatrix} c_{1131} & c_{1132} & c_{1133} \\ c_{2131} & c_{2132} & c_{2133} \\ c_{3131} & c_{3132} & c_{3133} \end{pmatrix}$$

let i=2,3 We can get similar $3 \times 3$ matrix $C_{ij}$ $i,j$=1,2,3. The elements of $C_{ij}$ are same as the formula in [3].

In the PML domain, apply the complex stretch operator to (10):

$$\sum_{j=1}^{3} \frac{\partial \tilde{\emptyset}}{\partial \tilde{x}_j}\left(C_{j1kl} : \frac{\partial \tilde{u}_k}{\partial \tilde{x}_l}\right) = \sum_{j=1}^{3} \frac{\partial \tilde{\emptyset}}{S_j \partial x_j}\left(C_{j1kl} : \frac{\partial \tilde{u}_k}{S_l \partial x_l}\right) = (\Lambda\nabla\tilde{\emptyset}) \cdot [C_{11}(\Lambda\nabla\tilde{u}_1) + C_{12}(\Lambda\nabla\tilde{u}_2) + C_{13}(\Lambda\nabla\tilde{u}_3)] \quad (11)$$

where

$$\Lambda = \begin{bmatrix} 1/S_1 & & \\ & 1/S_2 & \\ & & 1/S_3 \end{bmatrix} \quad (12)$$

Since $\Lambda$ is Diagonal, for each vector $\vec{a}$ and $\vec{b}$ :

$$(\Lambda\vec{a}) \cdot \vec{b} = \frac{a_1 b_1}{S_1} + \frac{a_2 b_2}{S_2} + \frac{a_3 b_3}{S_3} = \vec{a} \cdot (\Lambda\vec{b})$$

$$\sum_{j=1}^{3} \frac{\partial \tilde{\emptyset}}{\partial \tilde{x}_j}\left(C_{j1kl} : \frac{\partial \tilde{u}_k}{\partial \tilde{x}_l}\right) = (\nabla\tilde{\emptyset})\Lambda \cdot [C_{11}(\Lambda\nabla\tilde{u}_1) + C_{12}(\Lambda\nabla\tilde{u}_2) + C_{13}(\Lambda\nabla\tilde{u}_3)] \quad (13)$$

apply the associate law of matrix product
$$\Lambda C(\Lambda \vec{a}) = (\Lambda C \Lambda)\vec{a}$$

we get

$$\sum_{j=1}^{3} \frac{\partial \tilde{\emptyset}}{\partial \tilde{x}_j}\left(C_{j1kl} : \frac{\partial \tilde{u}_k}{\partial \tilde{x}_l}\right) = \nabla\tilde{\emptyset} \cdot [(\Lambda C_{11}\Lambda)(\nabla\tilde{u}_1) + (\Lambda C_{12}\Lambda)(\nabla\tilde{u}_2) + (\Lambda C_{13}\Lambda)(\nabla\tilde{u}_3)] \quad (14)$$

Compare to (10), we get transformed $\tilde{C}_{ij}$ in the PML domain:
$$\tilde{C}_{ij} = \Lambda C_{ij}\Lambda \quad (14)$$

or $\tilde{C}_{ijkl} = C_{ijkl} \frac{S_1 S_2 S_3}{S_k S_l} \quad (14.1)$

It's easy to verify that $\tilde{C}_{ijkl} = \tilde{C}_{ijlk}$ and $\tilde{C}_{ijkl} = \tilde{C}_{jikl}$. So $\tilde{\tau}_{ij} = \tilde{\tau}_{ji}$.